\documentclass[12pt,a4paper]{article}
\usepackage{amssymb}
\usepackage{amsfonts}
\usepackage[english]{babel}
\usepackage{graphicx}

\newtheorem{theorem}{Theorem}[section]

\newcommand{\bean}{\begin{eqnarray*}}
\newcommand{\eean}{\end{eqnarray*}}
\newcommand{\ba}{\begin{array}}
\newcommand{\ea}{\end{array}}
\newcommand{\be}{\begin{equation}}
\newcommand{\ee}{\end{equation}}
\newcommand{\bea}{\begin{eqnarray}}
\newcommand{\eea}{\end{eqnarray}}
\newcommand{\pa}{\partial}

\newcommand{\no}{\nonumber}

\newcommand{\wt}{\widetilde}

\begin{document}
\title
{Parity-Time Symmetric Solitons in the Complex KP Equation  
\author{
 Jen-Hsu Chang \\Graduate School of National Defense, \\
 National Defense University, \\
 Tauyuan City,  335009, Taiwan }

 }

\date{}

\maketitle

\begin{abstract}
One constructs  the parity-time symmetric solitons in the complex KP Equation using the totally non-negative Grassmannian. We obtain that every element in the totally non-negative orthogonal  Grassmannian corresponds to a  parity-time symmetric solitons solution. 
\end{abstract}
Keywords: PT-Symmetry, Grassmannian,  complex KP Equation, Solitons \\
2020 Mathematics Subject Classification: 15A15; 15A24; 35B34; 37K40
\newpage

\section{Introduction}
During the last three decades the amount of information transmitted via optical fiber communication systems
has increased enormously. High performance of modern communication lines is provided by optical
fibers, where the information is encoded and transmitted as a sequence of light pulses. The capacity problem of the optical fiber communication systems requires deep understanding of  the basic physical principles underlying the technical implementations as well as mathematical models describing the features of the systems \cite{al}. One of the most important of such models is based on the concept of soliton. Solitons are self-localized wave packets that can propagate along wave-guides preserving their shape and velocity, and exhibit particle-like collisions with each-other. When the optical soliton was theoretically predicted \cite{HasegawaTappert1973} and experimentally observed \cite{MollenauerStolenGordon1980}, there was a promising idea that it can be used as information carrier in optical fiber communication systems \cite{HasegawaKodama1995} due to its exceptional robustness against perturbations.

The idea of parity-time (PT)-symmetry \cite{ko, ss} in the context of a wide class of non-Hermitian Hamiltonians having entirely real eigen-spectra was proposed in the work of Bender and Boettcher \cite{bb}. The two fundamental discrete symmetries in physics
are given by the parity operator, P, defined as
\[ P\Psi (\vec{x}, t) = \Psi(- \vec {x}, t) \]
and by the time reversal operator, T , defined in Wigner's sense as \cite{wi}
\[ T \Psi (\vec{x}, t) = \Psi^*( \vec {x}, -t), \]
where $*$ means the complex conjugation. The operator T is antilinear, i.e.,
\[ T (\lambda \psi + \phi) = \lambda^* T \psi + T \phi, \]
for any vectors $ \psi, \phi $ and a complex number $\lambda $.
Additionally,
\be P^2=T^2=I, \quad [P, T]=0. \label{P} \ee
An operator $H$ is PT symmetric if
\[ [PT, H]=0.  \]
It is not difficult to see that if $H$ is a PT symmetric and $\psi $ is an eigenfunction with eigenvalue
$\lambda$ , then $PT \psi=\psi^{*}(-\vec {x}, -t)$ is also an eigenfunction of $H$ with eigenvalue
$\lambda^*$. A  connection between PT symmetry and reality of the spectrum was pointed out in \cite{bb}. To emphasize this connection, Bender and Boettcher  introduced the notion of unbroken PT symmetry. A PT -symmetric operator H is said to be unbroken if any eigenfunction of H is at the same time an eigenfunction of the PT operator. A PT -symmetric operator H is said to be broken if any eigenfunction of H is not at the same time an eigenfunction of the PT operator. For the unbroken case, the relation $ H \Psi = E \Psi $
implies the existence of $\lambda $  such that $ PT \Psi  = \lambda  \Psi $. From
(\ref{P}), i.e.,
\[ PTPT \Psi =\Psi= PT( \lambda  \Psi) = \lambda^* PT \Psi= \lambda \lambda ^* \Psi, \]
it follows that there exists a real constant $\theta $ such
that $ \lambda  = e^{i \theta} $, i.e., any eigenvalue of the PT operator is
a pure phase. When  H is  a PT symmetric operator and also is  unbroken, its eigenvalues will be real value. It can be seen as follows.  Suppose  $H \Psi= E\Psi$. Then one has
\[ H (PT \Psi)=H( e^{i \theta}  \Psi)=e^{i \theta} E \Psi= PT(H\Psi)= E^* PT \Psi=E^* e^{i \theta} \Psi. \]
Hence $E$ is real. But if H is  a PT symmetric operator and  broken, its eigenvalues will be complex value.

Let's consider  the Hamiltonian operator
\be  H = - \frac{d^2}{dx^2} -\frac{d^2}{dy^2}  + u (x,y ), \label{ha} \ee
where we have the complex potential $ u (x,y ) = P(x,y) + i Q(x,y)$.  For $H$ to be PT-symmetric operator, we see that
$V^*(-x, - y )=V(x, y)$.  Then  its real and imaginary parts, $P (x, y )$  and $Q (x, y )$,   respectively,  obey the following
relationships \cite{ko,ss}:
\be  P (-x, -y ) = P (x, y ),  \quad  Q (-x, -y ) = - Q (x, y ), \label{pt} \ee
which means that $P(x, y )$ is an even function and $Q(x, y )$ is an odd function. In the paraxial approximation in geometry optics, which assumes that the ray of light makes a small angle to the central
propagation direction $z$. Then we also have the Schr\"odinger equation
\be  i \Psi_z=H \Psi,  \label{sch} \ee
where $V(x, y)$ is the complex refractive index (or potential) and $\Psi$ is the amplitude of electric field (or wave function). If PT symmetric is unbroken, a medium with gain and loss allows for a stationary propagation of light, while the light undergoes attenuation or amplification if PT symmetric is broken. Hermitian Hamiltonians correspond to the cases where the optical energy is conserved and $V(x,y)$ is real. On the other hand, the presence of losses or gain in optical structures maps to non-conservative non-Hermitian operator (\ref{ha}) with a complex potential \cite{ba}. But if the Hamiltonian operator (\ref{ha}) is PT symmetric and unbroken, then it has real spectrum. In this case, there is a effective balance of gain and lose \cite{zy}. \\
\indent To construct PT-symmetric type line solitons, we consider the complex KP-(II) equation. 
\be    \pa_x (-4 u_t+u_{xxx}+6uu_x)+ 3u_{yy}=0,   \label{kp} \ee
where $u(x,y,t) $ is complex number and $x,y,t$ are real numbers. Let $u=P+iQ$,  and then we have the coupling system , 
\[
 \pa_x [ -4 P_t+P_{xxx}+6(PP_x-QQ_x)]+ 3P_{yy}=0,  \quad \pa_x [ -4 Q_t+Q_{xxx}+6(PQ)_x]+ 3Q_{yy}=0,
\]
where $P $ and $Q$ are real numbers. The resonant interaction of the real KP-(II) equation (\ref{kp}) plays a fundamental  role in multi-dimensional wave phenomenon. It has attracted much attractions using the totally non-negative Grassmannians \cite{bc, ko1, ko3}, that is, those points of the real Grassmannian whose Plucker coordinates are all non-negative.  For the KP-(II) equation case, the  $\tau$-function  is described by the Wroskian form with respect to $x$ obtained from the Hirota Bilinear form \cite{hi}. There are three basic types of interactions: the "X" -shaped P-type and  O-type solitons, and the "box"-shaped T-type soliton (the resonance) \cite{ko6}. When $u$ is independent of $y$, we have the complex KdV equation \cite{cc, cf1, cf2}
 \be u_{t}+ 6uu_x+u_{xxx}=0. \label{kd} \ee
The interaction of solitons  in the complex KdV equation (\ref{kd}) is of P-type.  \\
\indent The paper is organized as follows. In section 2, one constructs the real soliton solution using the $\tau$ function and  the totally non-negative Grassmannian in KP-(II) theory. And then the complex multi-line solitons  of the KP-II equation is introduced. In section 3, we investigate  the relation between the PT symmetric solitons and totally non-negative Orthogonal Grassmannian.  In section 4, we conclude the paper with several remarks.

\section{Complex KP Equation}
In this section, one constructs the multi-line solitons of the KP-(II) equation using the $\tau$-function. We express the $\tau$-functions as the linear combinations of the phase functions over the coefficients satisfying the Plucker relations. \\
\indent The complex KP-(II) equation (\ref{kp}) has the Hirota equation 
\be (-4D_t D_x +D_x^4+3  D_y^3) \tau_N \circ  \tau_N=0, \label{hr} \ee
where $u(x,y,t)=2\ln \pa_x^2 \ln \tau_N(x,y,t)$. To construct $\tau_N(x,y,t)$, we define  
\be
\tau_N= det 
\left[\ba{cccc} f_1^{(0)}  & f_1^{(1)} & \cdots   &  f_1^{(N-1)}    \\
 f_2^{(0)} &  f_2^{(1)} & \cdots   &   f_2^{(N-1)} \\
\vdots  & \vdots & \vdots & \vdots    \\
 f_N^{(0)}& f_N^{(1)} & \cdots  &   f_N^{(N-1)}   \ea \right], \label{hi}   \ee
where the elements in the above determinant are defined by ($ i=1,2,3 \cdots, N $ )
\be  
\frac{\pa f_i}{\pa x_m}  = \frac{\pa^m f_i}{ \pa x^m} , \quad x_1=x, \quad x_2=y, \quad x_3=t,  \label{lin} 
\ee
and $f_i^{(n)}$ means the n-th order derivative with respect to $x$, $n=0,1,2,3, \cdots, N-1$. Also, we can write $\tau_N $ as a Wronskian, i.e., $(f_i^{(0)}=f_i)$, 
\[ \tau_N=Wr( f_1^{(0)}, f_2^{(0)}, f_3^{(0)}, \cdots, f_N^{(0)} ). \]
It has been shown that (\ref{hi}) satisfies the Hirota equation (\ref{hr})\cite{hi}. \\
\indent Now, we construct the real resonant solutions of KP-(II) equation using the totally non-negative Grassmannian in KP-(II) theory \cite{bc, ko1, ko3}. Here one considers a finite dimensional solution
\bea 
f_i (x,y,t)  &=& \sum_{j=1}^M a_{ij} E_j (x,y,t), \quad i=1,2, \cdots N < M, \no \\
E_j (x,y,t) &= & e^{\eta_j}, \quad  \eta_j= k_j x+k_j^2 y+ k_j^3 t + \xi_j, \quad j=1,2, \cdots M \label{eq} 
\eea 
where $k_j $ is real number and $\xi_j$ are parameters. Each $E_j(x,y,t)$ satisfies the equations (\ref{lin}). Then each resonant solution of KP-(II) equation can be parametrized  by a full rank matrix 
\[ A= \left[\ba{cccc} a_{11}  & a_{12}  & \cdots   &  a_{1M}   \\
 a_{21} &  a_{22} & \cdots   &   a_{2M} \\
\vdots  & \vdots & \vdots & \vdots    \\
 a_{N1} & a_{N2}  & \cdots  &   a_{NM}    \ea \right] \in M_{N \times M} (\textbf{R}). \]  
Using the Binet-Cauchy formula, the $\tau$-function $\tau_N $ can be written as 
\bea 
\tau_A &=& \tau_N = Wr (f_1, f_2, \cdots, f_N)=det 
\left[\ba{cccc} f_1 & f_1^{'} & \cdots   &  f_1^{(N-1)}    \\
 f_2 &  f_2^{'} & \cdots   &   f_2^{(N-1)} \\
\vdots  & \vdots & \vdots & \vdots    \\
 f_N & f_N^{'} & \cdots  &   f_N^{(N-1)}   \ea \right]  \no \\
&=& det \left [\left(\ba{cccc} a_{11}  & a_{12}  & \cdots   &  a_{1M}   \\
 a_{21} &  a_{22} & \cdots   &   a_{2M} \\
\vdots  & \vdots & \vdots & \vdots    \\
 a_{N1} & a_{N2}  & \cdots  &   a_{NM}    \ea \right) \left(\ba{cccc} E_1  & k_1 E_1 & \cdots   &  k_1^{N-1}E_1   \\
 E_2 & k_2 E_2 & \cdots   &   k_2^{N-1} E_2 \\
\vdots  & \vdots & \vdots & \vdots    \\
 E_M & k_M E_M  & \cdots  &    k_M^{N-1} E_M   \ea \right)  \right] \no \\
&=& \sum_J \Delta_J (A) E_J (x,y,t), \label{ta}
\eea 
where $\Delta_J (A)$ is the  $ N \times N $ minor for the columns  with the index set $J=\{ j_1, j_2, j_3, \cdots, j_N\} $, and $E_J$ is the Wronskian 
\be  E_J= Wr (E_{j_1},E_{j_2}, E_{j_3}, \cdots, E_{j_N} )= \prod_{m< l} (k_{j_l}-k_{j_m}) E_{j_1} E_{j_2}E_{j_3} \cdots E_{j_N}. \label{va} \ee
We notice that the coefficients $ \Delta_J (A)$ of $\tau_A$  have to satisfy the Plucker relations. 
Also,  to make the product in (\ref{va}) is positive, one also assumes that  
\be     k_1< k_2  <k_3< \cdots  < k_N .   \label{pv} \ee
Each line soliton is obtained by the balance between adjacent regions and is localized only at the boundaries of the dominant regions. For any fixed time, suppose $ E_{i} E_{j_2}E_{j_3} \cdots E_{j_N}$ and $E_{j} E_{j_2}E_{j_3} \cdots E_{j_N} $ are adjacent regions. In general,  one has the boundary, i.e., the  line soliton $ [i,j]$-soliton. From (\ref{ta}) and (\ref{va}), it can be seen locally 
\bean  \tau_A  & \approx   & det (A_I) \prod_{j_l< j_m} (k_{j_l}-k_{j_m})  E_i E_{j_2} E_{ j_3}    \cdots, E_{j_N} \\
& + & det (A_J)\prod_{ j_l< j_m} (k_{j_l}-k_{j_m})    E_j E_{j_2} E_{ j_3}    \cdots, E_{j_N} ,   \eean 
where $I=\{i, j_2, j_3,  \cdots , j_N\} $ and $J =\{j , j_2, j_3,  \cdots , j_N\}$. Since $u=2 \pa_{xx} (\ln \tau_A)$ in (\ref{kp}), we have the  $ [i,j]$-soliton 
\bea  u &=& 2 \pa_{xx} [1+ e^{(k_i-k_j)x+ (k_i^2-k_j^2)y+  (k_i^3-p_j^3)t+ \mu_i-\mu_j} ] \no  \\ 
&=& \frac{(k_i-k_j)^2}{2} sech^2 \frac{(k_i-k_j)x+ (k_i^2-k_j^2)y+  (k_i^3-k_j^3)t+\mu_i-\mu_j }{2}
 \label{am}.
\eea
where 
\bean 
\mu_i &=& \eta_i+ \ln \left [ det (A_I) \prod_{j_l< j_m }   (k_{j_l}-k_{j_m})  \right ] \\
\mu_j &=&\eta_j +\ln  \left [det (A_J)\prod_{ j_l< j_m} (k_{j_l}-k_{j_m})  \right ].  
\eean 
Next, we assume $u=P(x,y,t)+ i Q(x,y,t)$ and satisfies the PT-symmetry condition (\ref{pt}) or 
\be u^*(-x, -y, -t)= u(x,y,t) \label{pt1}. \ee
To obtain solutions of complex KP-(II) equation, we consider the condition in (\ref{eq}) by 
\be  \xi_j= \mu_j+i \theta_j , \label {pe} \ee
where $\mu_j$ and $\theta_j$ are  real. Firstly, one considers a complex line soliton of the complex KP-(II) equation (\ref{kp}) and takes the $\tau$-function
\[\tau_1 (x,y,t)= f_1+f_2=e^{\eta_1}+e^{\eta_2}= e^{ k_1 x+k_1^2 y+ k_1^3 t + \mu + i \theta_1}+e^{ k_2 x+k_2^2 y+ k_2^3 t + \mu+ i \theta_2},   \]
where one uses the constraint $\mu_1=\mu_2=\mu $ to satisfy the PT-symmetric condition (\ref{pt1}). 
We notice that 
\[ \tau_1^* (-x,-y,-t)=e^{-\eta_1}+e^{-\eta_2}=e^{-\eta_1-\eta_2} (e^{\eta_1}+e^{\eta_2})=e^{-\eta_1-\eta_2} \tau_1 (x,y,t). \]
Then from $u(x,y,t)=2 \pa_{xx} \ln \tau_1 (x,y,t)=2 \pa_{xx} \ln \tau_1^*(-x,-y,-t) $ one can get 
\bea u^*(-x,-y,-t)&=& u(x,y,t)=\frac{A_{[1,2]}+ A_{[1,2]} cos (\theta_2-\theta_1) cosh \Delta_{[1,2]}}{ [cos (\theta_2-\theta_1)+ cosh \Delta_{[1,2]}]^2}\no \\
&+& i \frac{A_{[1,2]} sin (\theta_2-\theta_1) sinh \Delta_{[1,2]}}{ [cos (\theta_2-\theta_1)+ cosh \Delta_{[1,2]}]^2}=P(x,y,t)+ i Q(x,y,t),  \label{kp1} \eea 
where 
\[A_{[1,2]}=(k_2-k_1)^2, \quad \Delta_{[1,2]}= [(k_2-k_1)x+(k_2^2-k_1^2)y +   (k_2^3-k_1^3)t ]. \]
The maximum of line soliton of $P(x,y,t)$ is determined by 
$ \Delta_{[1,2]}=0$; and the minimums  of $P(x,y,t)$ are the lines 
\[ \Delta_{[1,2]}=\pm \frac{1}{k_2-k_1} arc cosh [\cos(\theta_2-\theta_1)-2 \sec (\theta_2-\theta_1)]. \]
 The minimum and maximum of $Q(x,y,t)$ is determine by the equation \cite{cc}
\[ \Delta_{[1,2]}=\pm \frac{1}{k_2-k_1} arccosh \left [\frac{1}{2} cos (\theta_2-\theta_1) +\frac{\sqrt{2}}{4} \sqrt{17+cos[2(\theta_2-\theta_1)]} \right ].  \]
Please see the figure 1.  When $\theta_1-\theta_2=0$, we get the real one-line soliton. And there is a singular line soliton when $\theta_1-\theta_2=\pi$.   Also, if $\theta_2-\theta_1=\pm \pi/2$,   
\[  u(x,y,t)=A_{[1,2]} \left [ sech^2 \Delta_{[1,2]} \pm  i \tanh \Delta_{[1,2]} sech \Delta_{[1,2]}\right ]. \]
This is a particular case of  the Scarff  II potential \cite{gm}. 
\begin{figure}
	\centering
		\includegraphics[width=1.0\textwidth]{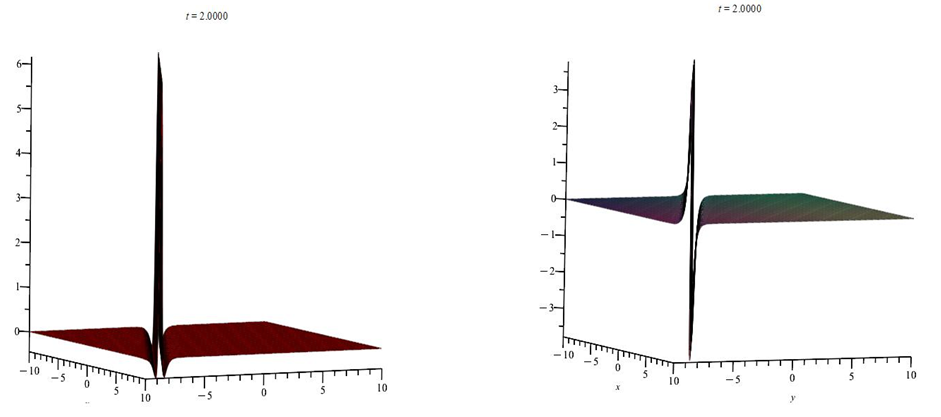}
		\caption{One line soliton: $k_1=0.1,  k_2=2, \theta_2-\theta_1=\frac{3 \pi}{4}. $   The left panel is the real part $P(x,y,t)$ and the right panel is the imaginary one $Q(x,y,t)$ in (\ref{kp1}). We notice the real part $P(x,y,t)$ has negative values, which is different from the real KP equation.   }
\end{figure}

\section{ PT symmetry and Orthogonal Grassmannian Manifold} 
In this section, one studies the relation between  PT symmetry and  Orthogonal Grassmannian. Given an Orthogonal Grannmannian, one can construct a PT symmetric multi-line solitons associated with it. \\
\indent To get the condition (\ref{pt1}), we see that $\tau$-function has the sufficient condition 
\be \tau^* (-x, -y, -t)= \tau (x,y,t). \label{ptt} \ee 
In such situations, we have to investigate the elastic solitons of KP-(II) equation \cite{bc} and the corresponding   self-dual $\tau$ -functions for the KP equation \cite{ko6, ck}. The elastic solitons means 
those  line solitons  for which the sets of incoming and outgoing  $ ( |y| \to \infty)  $ asymptotic line solitons are the same.  As mentioned before, there are three basic types of elastic solitons: P-type, O-type and T-type. The N-soliton solutions are combinations of these three line solitons \cite{ck}.  Notice that the "X" -shaped P-type and O-type  solitons are not resonant case, but the "box"-shaped T-type soliton is the resonant one. For the PT symmetric condition (\ref{ptt}), the Orthogonal Grassmannian Manifold is considered. \\
 \indent For N-soliton solutions, one considers an $ N \times 2N$ matrix. From the equation (\ref{ta}), $\tau$-function is a non-negative sum of the same exponential phase combinations  with a different sets of coefficients satisfying  the Plucker relations. Using (\ref{ta}), we see that 
\[ \tau(-x, -y, -t)= e^{-(\eta_1+\eta_2+\cdots+\eta_N)} \wt \tau(x,y,t), \] 
where 
\be \wt \tau(x,y,t)=\sum_{J \subset [2N],\quad \sharp J=N} det (A_J) K_JE_{i_1} E_{i_2}E_{i_3} \cdots E_{i_N} , \label{pr} \ee
where $K_J= \prod_{ j_l< j_m} (k_{j_m}-k_{j_l})$. Here $\{j_1, j_2, \cdots, j_N\}$ and $\{i_1, i_2, \cdots, i_N\}$ forms a disjoint partition of $\{1,2,3,4, \cdots, 2N\}$. Thus, we can define an Orthogonal Grassmannian manifold  \cite{gp, ko6}: given an $N \times 2N$, irreducible, rank $N$ coefficient matrix $A$ ,  all of its  $N \times N$ minors satisfy the duality conditions
\be det (A_J)= det (A_I),  \label{du} \ee
where $J =\{ j_1, j_2, \cdots, j_N   \}$ and $ I= \{i_1, i_2, \cdots, i_N\}$ forms a disjoint partition of $[2N]=\{1,2,3,4, \cdots, 2N\}$.   This condition (\ref{du}) can be proved to be equivalent to the condition \cite{jh, ko6}
\[ A D A^T=0, \]
where $A^T$ means the transpose of $A$ and $D=diag[-1,1,-1,1, \cdots , -1,1]$. 
Then  the Orthogonal Grassmannian manifold can also be defined by 
\be  OG(N, 2N)=\{ A \in R^{N\times 2N} |  A D A^T=0 \} , \label{og} \ee
It has dimension $\frac{N(N-1)}{2}$ and can be described as follows .There is an embedding from $N \times N $  symmetric matrix $M=[m_{i,j}]$ having $m_{i,i}=1$  to $A=[a_{i,j}] \in OG(N, 2N)$. The matrix $A$  is defined by 
\[  \left \{\begin{array}{ll} a_{i,2j-1}=a_{i, 2j}=m_{i, j}=1  \quad \mbox { if  $i =j$}                   \\
a_{i,2j-1}=-a_{i, 2j}=(-1)^{i+j+1} m_{i, j} &\quad  \mbox { if  $i <j$} \\
a_{i,2j-1}=-a_{i, 2j}=(-1)^{i+j}  m_{i, j} & \quad \mbox { if  $i >j$} \\
\end{array} \right. \]
For example, 
\[M=\left[\ba{cccc} 1 & m_{12}& m_{13}&m_{14}    \\ 
m_{12}  & 1& m_{23}&m_{24}  \\
 m_{13}& m_{23}& 1&m_{34}   \\  m_{14}& m_{24} & m_{34} &1 
\ea  \right]   \to \]
\[A=\left[\ba{cccccccc} 1 & 1& m_{12}& -m_{12}  &- m_{13}   &   m_{13}    & m_{14}   &  - m_{14}  \\ 
 -m_{12}& m_{12}  &1 &1 & m_{23}   &   -m_{23}    &- m_{24}   &   m_{24}  \\
  m_{13}&- m_{13}  &- m_{23}   &   m_{23}   &1 & 1  &m_{34}   &   -m_{34}  \\  
-m_{14}& m_{14}  & m_{24}   &   -m_{24}   &-m_{34}   &   m_{34}  &1&1 \\
\ea  \right] .   \]
Then we have the following 
\begin{theorem}\cite{gp}
If  the Orthogonal Grassmannian manifold $OG(N, 2N) $ is totally non-negative, then   it is  homeomorphic to an $N \choose 2 $-dimensional closed ball.
\end{theorem}
Thus, if $A$ is element in the  totally non-negative Orthogonal Grassmannian manifold, then the equation
 (\ref{pr})  can be written as 
\be \wt \tau(x,y,t)=\sum_{J \subset [2N],\quad \sharp J=N} det (A_I) K_J E_{i_1} E_{i_2}E_{i_3} \cdots E_{i_N} .\label{op} \ee
Next, let's choose $\mu _j$ in (\ref{pe}) defined in the phase $\eta_j$ as 
\be \mu_j= \frac{1}{2}  \sum_{r \neq j} \ln   |(p_r-p_j) |, \quad j, r =1,2,3, \cdots, 2N . \label{pa} \ee
Then we have 
\be  e^{ \sum_{m=1}^N \mu_{i_m}}= \prod_{m=1}^N \prod_{r \neq i_m}  |(p_r-p_{i_m}) |^{1/2}  =K_J^{\frac{-1}{2} } K_I^{\frac{1}{2}}  K_{[2N]}^{\frac{1}{2}} \quad   \mbox{with} \quad I \cup J=[2N],  \label{ff} \ee
where $K_{[2N]}= \prod_{i <j}  (p_i-p_j) ,$ an overall factor. Plugging (\ref{ff}) into (\ref{op}), using the duality condition (\ref{du}), one has 
\bea 
\wt \tau(x,y,t) &=& \sum_{J \subset [2N]} det (A_I) K_J^{\frac{1}{2} } K_I^{\frac{1}{2}}  K_{[N]}^{\frac{1}{2}}\hat E_{i_1} \hat  E_{i_2} \hat E_{i_3}  \cdots \hat E_{i_N}  \label{fi}  \\
&=& K_{[N]}^{\frac{1}{2}} \sum_{J  \subset [2N]} det (A_I) K_J^{\frac{1}{2} } K_I^{\frac{1}{2}} \left  ( \hat E_{i_1} \hat  E_{i_2} \cdots \hat E_{i_N} + \hat E_{j_1} \hat  E_{j_2}  \cdots \hat E_{j_N}\right) , \no
\eea
where $\hat E_m=  e^{ k_m x+k_m^2 y+ k_m^3 t + i \theta_m}$. Consequently, the $\tau$-function defined by (\ref{fi} ) satisfies the PT symmetric condition (\ref{ptt}). But singular solitons could happen due to the functions $ e^{i (\theta_{i_1}+\theta_{i_2}+\theta_{i_3}+ \cdots+ \theta_{i_N})  }$  and $ e^{i (\theta_{j_1}+\theta_{j_2}+\theta_{j_3}+ \cdots+ \theta_{j_N} ) }$. To obtain non-singular solitons, we can take $\theta_m \geq 0, m=1, 2, 3, \cdots, 2N $  and $\sum_{m=1}^{2N} \theta_m \leq   \frac{\pi}{2}. $  \\
\indent  For example, we consider the $2 \times 4$ totally non-negative Orthogonal Grassmannian manifold
\[ A=\left[\ba{cccc} 1 & 1 & m_{12} & -m_{12}   \\  -m_{12} & m_{12}   & 1 & 1  \ea
 \right], \]
where  the one dimensional ball is $0 \leq m_{12} \leq 1$. For PT symmetry, we choose the parameters by (\ref{pa})
\bean
\mu_1 &=& \frac{1}{2} [\ln (k_2-k_1)+\ln (k_3-k_1) +\ln (k_4-k_1)] \\
\mu_2 &=& \frac{1}{2} [\ln (k_2-k_1)+\ln (k_3-k_2) +\ln (k_4-k_2)]  \\
\mu_3 &=& \frac{1}{2} [\ln (k_3-k_1)+\ln (k_3-k_2) +\ln (k_4-k_3)]  \\
\mu_4 &=& \frac{1}{2} [\ln (k_4-k_1)+\ln (k_4-k_2) +\ln (k_4-k_3)]
\eean 

Using (\ref{ta}) and (\ref{fi}), one has 
\bea
&&\tau_A=W(f_1,f_2, f_3, f_4) = \sqrt{(k_2-k_1) (k_3-k_1)(k_4-k_1)(k_3-k_2)  (k_4-k_2) (k_4-k_3) } \no \\
&&\left [2m_{12} \sqrt{(k_2-k_1) (k_4-k_3) } (\hat E_1 \hat E_2+ \hat E_3 \hat E_4) 
+ (1+m_{12}^2)  \sqrt{(k_3-k_1) (k_4-k_2) } (\hat E_1 \hat E_3+ \hat E_2 \hat E_4)\right. \no \\
&& + \left. (1-m_{12}^2)  \sqrt{(k_4-k_1) (k_3-k_2) } (\hat E_1 \hat E_4+ \hat E_2 \hat E_3) \right] \no \\
&& \equiv \left [2m_{12} \sqrt{(k_2-k_1) (k_4-k_3) } (\hat E_1 \hat E_2+ \hat E_3 \hat E_4) 
+ (1+m_{12}^2)  \sqrt{(k_3-k_1) (k_4-k_2) } (\hat E_1 \hat E_3+ \hat E_2 \hat E_4)\right. \no  \\
&& + \left. (1-m_{12}^2)  \sqrt{(k_4-k_1) (k_3-k_2) } (\hat E_1 \hat E_4+ \hat E_2 \hat E_3) \right] \no \\
&& \equiv \tau_A^*(-x,-y, -t).  \label{to}
 \eea
We have three basic types of solitons. 
\begin{itemize}
\item O-type soliton ($m_{12}=0$): \\ 
 \[ A_O=\left[\ba{cccc} 1 & 1 & 0 & 0  \\ 0 & 0 & 1 & 1  \ea
 \right], \]
 All the line solitons are symmetric with respect to $(x,y)=(0,0)$ when  $t=0$. Then 
\[ \tau_O \equiv  \cosh \Theta_O^+ + \sqrt{\Delta_O} \cosh \Theta_O^-, \]
where 
\bean  
\Theta_O^{\pm}&=&\frac{1}{2} [(\eta_2-\eta_1) \pm (\eta_4-\eta_3)], \quad \eta_j=x k_j+ y k_j^2+ t k_j^3 +i \theta_j, \quad j=1,2,3,4, \\
\Delta_O &=& \frac{(k_4-k_1)(k_3-k_2)}{(k_4-k_2)(k_3-k_1)}. 
\eean
 \item P-type soliton ($m_{12}=1$): \\
 \[ A_P=\left[\ba{cccc} 1 & 1 & 1& -1  \\ -1 & 1 & 1 & 1 \ea
 \right], \]
Similarly, all the line solitons are symmetric with respect to $(x,y)=(0,0)$ when  $t=0$. Then 
\[ \tau_P \equiv \sqrt{\Delta_P}  \cosh \Theta_P^+ + \cosh \Theta_P^-, \]
where 
\bean  
\Theta_P^{\pm}&=&\frac{1}{2} [(\eta_4-\eta_1) \pm (\eta_3-\eta_2)], \quad \eta_j=x k_j+ y k_j^2+ t k_j^3 +i \theta_j, \quad j=1,2,3,4, \\
\Delta_P &=& \frac{(k_4-k_3)(k_2-k_1)}{(k_4-k_2)(k_3-k_1)}. 
\eean
\item T-type soliton ($  0<m_{12}<1$): The equation (\ref{to}) can be written as 
\bea  \tau_T  &\equiv&   \left [2m_{12} \sqrt{\Delta_P}   \cosh \Theta_P^ +  + (1+m_{12}^2)  \cosh \Theta_O^+   +  (1-m_{12}^2)  \sqrt{\Delta_O}  \cosh \Theta_O^{-} \right ] \no \\
&=& \left [ 2m_{12}\sqrt{\Delta_P}  \cosh \hat \Theta_P^ + \cos (\theta_4+ \theta_3-\theta_1-\theta_2)\right.  \no \\
& +& \left. (1+ m_{12}^2)\cosh \hat \Theta_O^ +  \cos (\theta_2+ \theta_4-\theta_1-\theta_3) \right. \no \\
&+ &\left. (1- m_{12}^2)  \sqrt{\Delta_O} \cosh \hat \Theta_O^ -  \cos (\theta_2+ \theta_3-\theta_1-\theta_4) \right] \no \\
&+& i \left [  2m_{12}\sqrt{\Delta_P}  \sinh \hat \Theta_P^ + \sin (\theta_4+ \theta_3-\theta_1-\theta_2) \right. \no \\
& +&\left.   (1+ m_{12}^2)\sinh \hat \Theta_O^ +  \sin (\theta_2+ \theta_4-\theta_1-\theta_3) \right. \no \\
&+ & \left. (1- m_{12}^2) \sqrt{\Delta_O}  \sinh \hat \Theta_O^ -  \sin (\theta_2+ \theta_3-\theta_1-\theta_4),  \right]  \label{td}  \eea 
where 
\bean
\hat \Theta_P^{+}&=&\frac{1}{2} [(\hat \eta_4-\hat \eta_1) + (\hat \eta_3-\hat \eta_2)], \quad 
\hat \Theta_O^{\pm}=\frac{1}{2} [(\hat \eta_2-\hat \eta_1) \pm (\hat \eta_4-\hat \eta_3)] , \\
&& \hat  \eta_j=x k_j+ y k_j^2+ t k_j^3 , \quad j=1,2,3,4. \\
\eean 
\end{itemize}
From $\tau_T$, we see that singular solitons could happen. To get non-singular solitons, one could choose $ 0 \leq  \theta_1 \leq \theta_2 \leq  \theta_3 \leq  \theta_4 \leq \frac{\pi}{2}$ appropriately. Please see the figure 2.
\begin{figure}[t]
	\centering
		\includegraphics[width=0.90\textwidth]{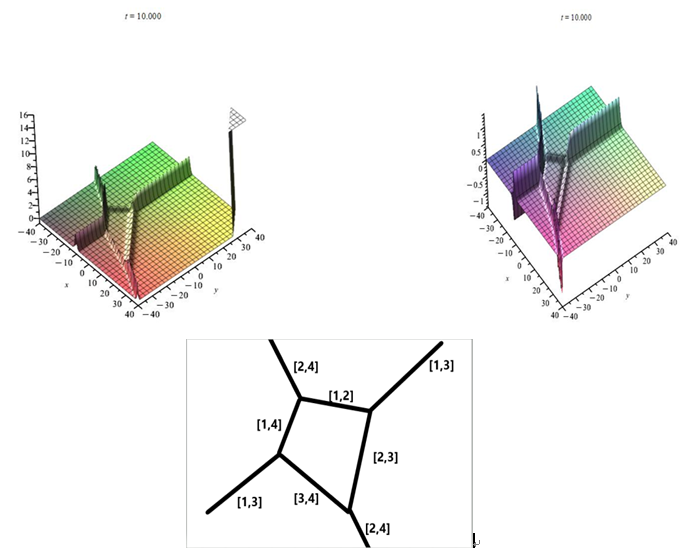}
\caption{ T-type resonant soliton:  $k_1=-1, k_2=0, k_3=1,k_4=2 ,  \theta_1=\pi/4, \theta_2=\pi/7, \theta_3=\pi/10, \theta_4=\pi/3, m_{12}=1/2. $  The left panel is the real part $P(x,y,t)$ and the right panel is the imaginary one $Q(x,y,t) $ using the $\tau$-function (\ref{td}). The bottom panel is the corresponding soliton graph.  }
\end{figure}

\section{Concluding Remarks } 
One constructs the PT symmetric solitons of the complex KP equation via the totally non-negative Orthogonal Grassmannian manifold. Given an $N \times 2N$  totally non-negative Orthogonal Grassmannian manifold, we have $ \frac{N(N-1)}{2} +N$ parameters for the the PT symmetric solitons. The PT symmetric solitons  are   non-singular or singular depending the choices of the parameters. We notice that  the non-singular soliton could be negative for the real part, which is different from the real KP equation; moreover, the imaginary one  could have a similar structure as the real part but  has  different amplitude. \\
\indent There are several questions to be studied. In \cite{cf1, cf2}, the real value energy and the regularized degenerate multi-solitons are investigated for the PT-symmetric solitons (P-type ) of the KdV equation. Here the degenerate means some parameters $k_j$ may take the same and the parameters $\theta_m$ can choose appropriately such that one obtains non-singular solitons. These can be consider similarly for the PT-symmetric  solitons of the complex KP equation. Also, the PT-symmetric solitons could be used as the potential of the Hamiltonian operator (\ref{ha}). Then we have $ \frac{N(N-1)}{2} +N+1$ parameters for this potential.  
It is known that a PT -symmetric operator H is said to be broken if any eigenfunction of H is not an eigenfunction of the PT operator. Thus, a broken PT-symmetry could happen for the PT-symmetric  solitons of the complex KP equation. Finally, the $\tau$-function in the KP theory can be used to construct  solitons solution in the Modified KP equation \cite{hi, ko6}. Then the Wadati potential for the PT symmetry  in the Modified KdV equation \cite{ba, wa} could be generalized to the Modified KP equation by the PT symmetric solitons (\ref{fi}). These issues need further investigations.

\subsection*{Acknowledgments}
This work is supported in part by the National Science  and Technology Council of Taiwan under
Grant No. NSC 111-2115-M-606-001.


\begin{thebibliography}{99}
\footnotesize
\bibitem{al} S. M. Alamoudi, U. Al Khawaja, and B. B. Baizakov, Averaged dynamics of soliton molecules in dispersion-managed optical fibers, Phys. Rev. A 89, 053817 (2014)
\bibitem{bc}G. Biondini and S. Chakravarty, “Soliton solutions of the Kadomtsev-Petviashvili II equation”, J. Math. Phys. 47, 033514 (2006) 

\bibitem{ba}I. V. Barashenkov, D. A. Zezyulin, and V. V. Konotop, Exactly solvable Wadati potentials in the PT-symmetric Gross-Pitaevskii equation, Pseudo-Hermitian Hamiltonians in Quantum Physics, May 18-23 2015, Palermo, Italy (Springer Proceedings in Physics, 2016),  arXiv:1511.06633
\bibitem{bb}C.M. Bender and S. Boettcher, Real spectra in non-hermitian hamiltonians having PT symmetry, Phys. Rev. Lett. 80, 5243–5246 (1998).
\bibitem{jh}Jen-Hsu Chang, Real Line Solitons  of  the BKP Equation, arXiv:2303.02385
\bibitem{cc}Julia Cen, Francisco Correa, Andreas Fring, Time-delay and reality conditions for complex solitons, Journal of Mathematical Physics 58, 032901 (2017)
\bibitem{cf1} Francisco Correa, Andreas Fring, Regularized degenerate multi-solitons, J. High Energ. Phys. (2016) 2016: 8
\bibitem{cf2}Julia Cen, Andreas Fring, Complex solitons with real energies,  J. Phys. A: Math. and Theor. 49 (2016) 365202
\bibitem{gm} Levai G. and  Znojil M., The interplay of supersymmetry and PT symmetry in quantum mechanics: a case study for the Scarf II potential, J. Phys. A: Math. Gen. 35 (2002), 8793-8804
\bibitem{gp} Pavel Galashin and  Pavlo Pylyavskyy, Ising model and the positive orthogonal Grassmannian, Duke Math. J. 169(10): p.1877-p.1942 (2020)
\bibitem{ko} Vladimir V. Konotop, Jianke Yang, Dmitry A. Zezyulin, Nonlinear waves in $\cal PT$-symmetric systems,  Reviews of Modern Physics 88, 035002 (2016)
\bibitem{hi} R. Hirota, The Direct Method in Soliton Theory, Cambridge Univ. Press, 2004
\bibitem{HasegawaTappert1973}
A. Hasegawa and  F. Tappert, Transmission of stationary nonlinear optical pulses in dispersive dielectric fibers ( I. )Anomalous dispersion, Appl. Phys. Lett., {\bf 42} (1973) 142--144
\bibitem{HasegawaKodama1995}
A. Hasegawa, Y. Kodama. {\it Solitons in optical communications}. Clarendon Press, Oxford, 1995
\bibitem{ko1}Yuji Kodama, KP solitons in shallow water, arXiv:1004.4607, 2010
\bibitem{ko3}Yuji Kodama, Lauren Williams,  KP solitons and total positivity for the Grassmannian
 arXiv:1106.0023, 2011
\bibitem{ko6} Y. Kodama, KP Solitons and the Grassmannians:  Combinatorics and Geometry of Two-Dimensional Wave Patterns, Springer Briefs in Mathematical Physics, Springer Nature Singapore, 2017
\bibitem{ck}S. Chakravarty and Y. Kodama, A generating function for the N-soliton solutions of the Kadomtsev-Petviashvili II equation, Contemp. Math., 471 (2008) 47-67.
\bibitem{MollenauerStolenGordon1980} L.F. Mollenauer, R.H.  Stolen, J.P.  Gordon.
Experimental observation of pico-second pulse narrowing and solitons in optical fiber.
Phys. Rev. Lett. {\bf 45} (1980) 1095-1098
\bibitem{ss} Sergey V. Suchkov, Andrey A. Sukhorukov, Jiahao Huang, Sergey V. Dmitriev, Chaohong Lee, Yuri S. Kivshar, Nonlinear switching and solitons in PT-symmetric photonic systems, Laser Photonics Rev. 10, 177 (2016)
\bibitem{wa} M Wadati, Construction of parity-time symmetric potential through the soliton theory. J Phys Soc Jpn 77, 074005, 2008
\bibitem{wi}E.Wigner, Group Theory and Its Application to Quantum Mechanics of Atomic Spectra, 1959 (New York: Academic).
\bibitem{zy}A A Zyablovsky, A P Vinogradov, A A Pukhov, A V Dorofeenko and  A A Lisyansky, PT-symmetry in optics, Physics Uspekhi 57 (11) 1063 - 1082 (2014)



\end{thebibliography}
\end{document}